\documentclass[conference]{IEEEtran}
\makeatletter
\def\ps@headings{%
\def\@oddhead{\mbox{}\scriptsize\rightmark \hfil \thepage}%
\def\@evenhead{\scriptsize\thepage \hfil \leftmark\mbox{}}%
\def\@oddfoot{}%
\def\@evenfoot{}}
\makeatother
\newif\ifnewtext
\newtexttrue		% uncomment to use new text

\usepackage{ifpdf}
\usepackage{amssymb}
\usepackage{color}
\usepackage{graphicx}
\usepackage[cmex10]{amsmath}
\usepackage{url}

\usepackage{algorithm}
\usepackage{algorithmic}
\usepackage{multirow}

\usepackage{subfigure}

\newcommand{\comment}[1]{}
\usepackage{xspace}

\newcommand{\ourproposal}{TOFEC\xspace}
\newcommand{\ourscheme}{TOFEC\xspace}

\newcommand{\normArrival}{\overline{\lambda}}
\newcommand{\aveUsage}{\overline{U}}
\newcommand{\fixedDelta}{\overline{\Delta}}
\newcommand{\linearDelta}{\widetilde{\Delta}}
\newcommand{\fixedExp}{\overline{\Psi}}
\newcommand{\linearExp}{\widetilde{\Psi}}

\newcommand{\nthreshold}{H^{N}}
\newcommand{\kthreshold}{H^{K}}
\newtheorem{theorem}{\textbf{Theorem}}

\newtheorem{corollary}{\textbf{Corollary}}

\newcommand{\onewidth}{0.6\columnwidth}
\newcommand{\twowidth}{0.45\columnwidth}

\newcommand{\fourwidth}{0.22\textwidth}

\newcommand{\shrinkbeforecaption}{-8pt}
\newcommand{\shrinkaftercaption}{-18pt}

\newcommand{\mybox}[1]{\vspace{5pt}\centerline{\framebox{\parbox[c]{0.95\columnwidth}{#1}}}\vspace{5pt}}

\begin{document}

\title{\ourproposal: Achieving Optimal Throughput-Delay Trade-off of Cloud Storage Using Erasure Codes}

\author{\IEEEauthorblockN{Guanfeng Liang and Ula\c{s}~C.~Kozat}
\IEEEauthorblockA{DOCOMO Innovations, Inc.,
Palo Alto, CA 94304\\
Email: \{gliang,kozat\}@docomoinnovations.com}
%\author{Guanfeng~Liang,~\IEEEmembership{Member,~IEEE,}
%        Shengbo~Chen,~\IEEEmembership{Student member,~IEEE,}
%        and~Ula\c{s}~C.~Kozat,~\IEEEmembership{Senior Member,~IEEE}
%\thanks{G. Liang and U.C. Kozat are with DOCOMO Innovations Inc., Palo Alto, California USA. G. Liang is the contact author. E-mail: gliang@docomoinnovations.com}
}
\maketitle

\begin{abstract}
Our paper presents solutions using erasure coding, parallel connections to storage cloud and limited chunking (i.e., dividing the object into a few smaller segments) together to significantly improve the delay performance of uploading and downloading data in and out of cloud storage.
\comment{
% We first focus on measuring the delay performance of a very popular cloud storage service Amazon S3. 
%We establish that there is significant randomness and independence  in service times for reading and writing different objects, as well as different segment of the same objects. 
%We demonstrate that 
}

%However, chunking and erasure coding impose extra overhead which reduces the number of requests the system can process in a given time period.

\ourscheme is a strategy that helps front-end proxy adapt to level of workload by treating scalable cloud storage (e.g. Amazon S3) as a shared resource requiring admission control. Under light workloads, \ourscheme creates more smaller chunks and uses more parallel connections per file, minimizing service delay. Under heavy workloads, \ourscheme automatically reduces the level of chunking (fewer chunks with increased size) and uses fewer parallel connections to reduce overhead, resulting in higher throughput and preventing queueing delay. Our trace-driven simulation results show that \ourscheme's adaptation mechanism converges to an appropriate code that provides the optimal delay-throughput trade-off without reducing system capacity. Compared to a non-adaptive strategy optimized for throughput, \ourscheme delivers $2.5\times$ lower latency under light workloads; compared to a non-adaptive strategy optimized for latency, \ourscheme can scale to support over $3\times$ as many requests.

\end{abstract}


\IEEEpeerreviewmaketitle

\section{Introduction}
\label{sec:intro}

Cloud storage has been gaining popularity rapidly as an economic, flexible and reliable data storage service that many cloud-based applications nowadays are implemented on. 
Typical cloud storage systems are implemented as key-value stores in which data objects are stored and retrieved via their unique keys. To provide high degree of availability, scalability, and data durability, each object is replicated several times within the internal distributed file system and sometimes also further protected by erasure codes to more efficiently use the storage capacity while attaining very high durability guarantees \cite{Huang12}. 

Cloud storage providers usually  implement a variety of optimization mechanisms such as load balancing and caching/prefetching internally to improve performance. Despite all such efforts, still evaluations of large scale systems indicate that there is a high degree of randomness in delay performance \cite{Garfinkel07anevaluation}. 
Thus, services that require more robust and predictable Quality of Service (QoS) must deploy their own external solutions such as sending multiple/redundant requests (in parallel or sequentially), chunking large objects into smaller ones and read/write each chunk through parallel connections, replicate the same object using multiple distinct keys, etc.

In this paper, we present \ourproposal ~-- a strategy that can provide much better throughput-delay performance for file accessing on cloud storage utilizing erasure coding. Although we base our analysis and evaluation on Amazon S3 service and present \ourproposal as an external solution, \ourproposal can be applied to many other cloud storage systems both externally and internally  with small modifications.

%Storage provider also monitors the load on each storage node and employs dynamic load balancing to prevent hot storage nodes that might observe high loads or slow nodes that have excessively high response times. Although mainly used for repairing data in unavailable storage nodes, some cloud providers also access coded blocks in parallel to uncoded blocks when uncoded blocks are stored in slow nodes \cite{Huang12}. 

\begin{figure}[!t]
\centering
\includegraphics[width = \onewidth]{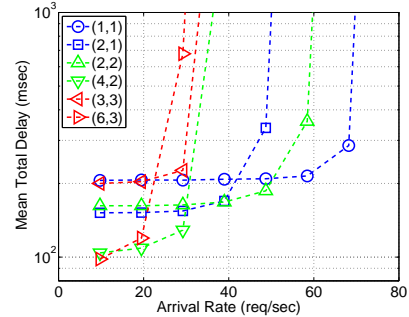}
\vspace{-7pt}
\caption{Delay for downloading 3MB files using fixed MDS codes}
\label{fig:fixedDelays}
\vspace{\shrinkaftercaption}
\end{figure}

\subsection{State of the Art}
Among the vast amount of research on improving cloud storage system's delay performance emerged in the past few years, two groups are in particular closely related to our work presented in this paper:

{\bf Erasure Coding with Redundant Requests:} As proposed by authors of \cite{fastcloud,Longbocodeingincloud, MDS-queue}, files are divided into a {\em pre-determined} number of $k$ chunks, each of which is $1/k$ the size of the original file, and encoded into $n>k$ of ``coded chunks'' using a $(n,k)$ Forward Error Correction (FEC) code, or more generally an Maximum Distance Separable (MDS) code. Downloading/uploading of the original file is accomplished by  downloading/uploading $n$ coded chunks using parallel connections simultaneously and is deemed served when download/upload of any $k$ coded chunks complete. Such mechanisms significantly improves the delay performance under light workload. However, as shown in our previous work \cite{fastcloud} and later reconfirmed by \cite{MDS-queue}, system capacity is reduced due to the overhead for using smaller chunks and redundant requests. This phenomenon is illustrated in Fig.\ref{fig:fixedDelays} where we plot the delay-throughput trade-off for using different MDS codes from our simulations using delays traces collected on Amazon S3. Codes with different $k$ are grouped in different colors. Using a code with high level of chunking and redundancy, in this case a $(6,3)$ code, although delivers $2\times$ gain in delay at light workload, reduces system capacity to only $30\%$ of the original basic strategy without chunking and redundancy, i.e., $(1,1)$ code!

This problem is partially addressed in \cite{fastcloud} where we present strategies that adjust $n$ according to workload level so that it achieves the near-optimal throughput-delay trade-off for the {\em predetermined} $k$. For example, if $k=3$ is used, the strategies in \cite{fastcloud} will achieve the lower-envelop of the red curves in Fig.\ref{fig:fixedDelays}. Yet, it still suffers from an almost 60\% loss in system capacity.

{\bf Dynamic Job Sizing:}
It has been observed in \cite{Garfinkel07anevaluation, stout} that in key-value storage systems such as Amazon S3 and Microsoft's Azure Storage, throughput is dramatically higher when they receive a small number of storage access requests for large jobs (or objects) than if they receive a large number of requests for small jobs (or objects), because each storage request incurs overheads such as networking delay, protocol-processing, lock acquisitions, transaction log commits, etc. Authors of \cite{stout} developed Stout in which requests are dynamically batched to improve throughput-delay trade-off of key-value storage systems. Based on the observed congestion Stout increase or reduce the batching size. Thus, at high congestion, a larger batch size is used to improve the throughput while at low congestion a smaller batch size is adopted to reduce the delay.

\subsection{Main Contribution}
We introduce an adaptive strategy for accessing cloud storage systems via erasure coding, call \ourproposal (Throughput Optimal FEC Cloud), that implements dynamic adjustment of chunking and redundancy levels to provide the optimal throughput-delay trade-off. In other words, \ourproposal achieves the lower envelop of curves in all colors in Fig.\ref{fig:fixedDelays}.

The primary novelty of \ourproposal is its backlog-based adaptive algorithm for dynamically adjusting the chunk size as well as the number of redundant requests issued to fulfill storage access requests. This algorithm of variable chunk sizing can be viewed as a novel integration of prior observations from the two bodies of works discussed above. Based on the observed backlog level as an indicator of the workload, \ourproposal increases or reduces the chunk size, as well as the number of redundant requests. In our trace-driven simulation evaluation, we demonstrate that: (1) \ourproposal successfully adapt to full range of workloads, delivering $3\times$ lower average delay than the basic static strategy without chunking under light workloads, and under heavy workloads over $3\times$ the throughput of a static strategy with high chunking and redundancy levels optimized for service delay; and (2) \ourproposal provides good QoS guarantees as it delivers low delay variations.

\ourproposal works without any explicit information from the back-end cloud storage implementation: its adaptation strategy is implemented solely at the front-end application server (the storage client) and is based exclusively on the measured latency from unmodified cloud storage systems. This allows \ourproposal to be more easily deployed, as individual cloud applications can adopt \ourproposal without being tied-up with any particular cloud storage system, as long as a small number of APIs are provided by the storage system.
% that are common in most existing popular cloud storage systems such as Amazon S3 and Microsoft's Azure Storage.

\section{System Models}
\label{sec:system}

\begin{figure}[!t]
\centering
\includegraphics[width = 0.9\columnwidth]{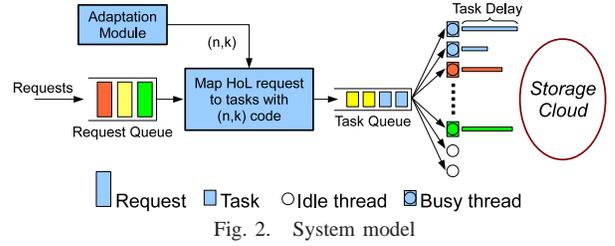}
\vspace{-10pt}
\caption{System model}
\label{fig:system}
%\vspace{\shrinkaftercaption}
\end{figure}

\subsection{Basic Architecture and Functionality}
The basic system architecture of \ourproposal captures how web services today utilize public or private storage clouds.  The architecture consists of proxy servers in the front-end and a key-value store, referred to as  storage cloud, in the back-end. 
%\ourproposal facilitates high-performance storage access for these services by controlling and adapting the way the proxy servers makes use of the cloud storage to provide the best possible response time for uploading or downloading files.
Users interact with the proxy through a high-level API and/or user interfaces.  
The proxy translates every high-level user request (to read or write a file) into a set of $n \ge 1$ tasks. Each task is essentially a basic storage access operation such as {\tt put, get, delete,} etc. that will be accomplished using  low-level APIs provided by the storage cloud. The proxy maintains a certain number of parallel connections to the storage cloud and each task is executed over one of these connections. After a certain number of tasks are completed successfully, the user request is considered accomplished and the proxy responds to the user with an acknowledgment. The solutions we present are deployed on the proxy server side transparent to the storage cloud.

For read request, we assume the file is pre-coded into $n^{max}\ge n$ coded chunks with an $(n^{max},k)$ MDS code and stored on the cloud. Completion of downloading any $k$ coded chunks provides sufficient data to reconstruct the requested file. For write request, the file to be uploaded is divided and encoded into $n$ coded chunks using an $(n,k)$ MDS code and hence completion of uploading any $k$ coded chunks means sufficient data have been stored onto the cloud. Thus, upon completion of a request, the $n-k$ un-started and/or unfinished tasks are then preemptively canceled and removed from the system.\footnote{For write request, the remaining tasks can also be scheduled as background jobs depending on the subsequent read profile of the file.}

Accordingly, we model the proxy by the queueing system shown in Fig.\ref{fig:system}. There are two FIFO (first-in-first-out) queues: (i) the {\em request queue} that buffers all incoming user requests, and (ii) the {\em task queue} that is a multi-server queue and holds all tasks waiting to being executed. $L$ threads\footnote{We avoid the term ``server'' that is commonly used in queueing theory literature to prevent confusion.}, representing the set of parallel connections to the storage cloud, are attached to the task queue. 
The adaptation module of \ourproposal monitors the state of the queues and the threads, and decides what coding parameter $(n,k)$ to be used for each request. 
Without loss of generality, we assume that the head-of-line (HoL) request leave the request queue only when there is at least one idle thread {\bf and} the task queue is empty. A batch of $n$ tasks is then created for that request and injected into the task queue. As soon as any $k$ tasks complete successfully, the request is considered completed.  Such a queue system is work conserving since no thread is left idle as long as there is any request or task pending.

\comment{

Client requests arrive at any of the proxy servers. 
When client wants to upload a file, proxy server divides the file into one or more chunks, and for every chunk it creates one write task using appropriate storage cloud API's to carry out the actual uploading. When all write tasks complete successfully, the job is completed and a response is sent back to the client.  When client wants to download a file, proxy server checks which chunks need to be fetched from the storage cloud. Proxy generates read tasks for these chunks using appropriate API. After all read tasks complete successfully, the job is completed and the file is streamed back to the client. The solutions we present are deployed on the proxy server side transparent to the storage cloud.

Cloud storage has two main purposes: (1) Provide data storage with high durability and availability. (2) Provide on demand scaling of storage needs. Cloud storage does not interpret the objects it stores, but rather treats them as byte strings with a well-defined length. For high durability and availability, typical cloud systems replicate each object several times in different physical locations and may use FEC internally. From proxy servers', as well as \ourproposal's, perspective, cloud storage is a black box whose internal techniques are unknown. Proxy servers only know the response times for each query (e.g., putting, getting, copying, deleting objects) it sends to the cloud storage.
}

\subsection{Basics of Erasure Codes}
\label{ssec:model:code}
An $(n,k)$ MDS code (e.g., Reed-Soloman codes) encodes $k$ data chunks each of $B$ bits into a codeword consisting of $n$ $B$-bit long coded chunks. The coded chunks can sustain up to $n-k$ erasures such that the $k$ original data chunks can be efficiently reconstructed from {\bf any} subset of $k$ coded chunks.
$n$ and $k$ are called the length and dimension of the MDS code. We also define $r = n/k$ as the redundancy ratio of an $(n,k)$ MDS code. This erasure resistant property of MDS codes has been utilized in prior works \cite{fastcloud,Longbocodeingincloud, MDS-queue}, as well as in this paper, to improve delay of cloud storage systems. Essentially a coded chunk experiencing long delay is treated as an erasure. 

\begin{figure}[t]
\centering
\includegraphics[width = 0.9\columnwidth]{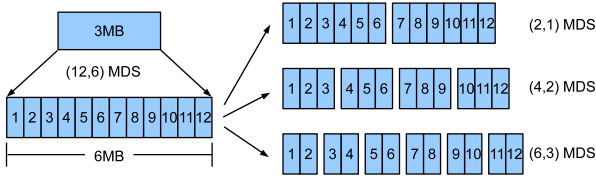}
\vspace{-5pt}
\caption{Example of supporting multiple chunk sizes with Shared Key approach: the 3MB file is divided and encoded into a coded file of 6MB consisting 12 strips, each of 0.5MB. Download the file using a $(2,1)$ MDS code is accomplished by creating two read tasks: one for strips 1-6, and the other for strips 7-12.} 
\label{fig:partialRead}
\vspace{\shrinkaftercaption}
\end{figure}

In this paper, we make use of another interesting property of MDS codes to implement variable chunk sizing of \ourproposal in a storage efficient manner:  MDS codes of high length and dimension for small chunk sizes can be used as MDS codes of smaller code length and dimension of larger chunk sizes. To be more specific, consider any $(N,K)$ MDS code for chunks of $b$ bits. To avoid confusion, we will refer to these $b$-bit chunks as strips. A different MDS code of length $n = N/m$, dimension $k=K/m$ and chunk size $B=b m$ for some $m>1$ can be constructed by simply batching every $m$ data/coded strips into one data/coded chunk. The resulting code is an $(n,k)$ MDS code for $B$-bit chunks because any $k$ coded chunks covers $mk = K$ coded strips, which is sufficient to reconstruct the original file of $Bk = b m \times K/m = bK$ bits.  This property is illustrated as an example in Fig. \ref{fig:partialRead}. In this example, a 3MB file is divided into 6 strips of 0.5MB and encoded into 12 coded strips of total size 6MB, using a $(12,6)$ MDS code. This code can then be used as a $(2,1)$ code for 3MB chunks, a $(4,2)$ code for 1.5MB chunks and a $(6,3)$ code for 1MB chunks {\bf simultaneously} by batching 6, 3 and 2 strips into a chunk.

\subsection{Definitions of Different Delays}
The delay experienced by a user request consists of two components: {\em queueing delay ($D_q$)} and {\em service delay ($D_s$)}. Both are defined with respect to the request queue: (i) the queueing delay is the amount of time a request spends waiting in the request queue and (ii) the service delay is the period of time between when the request leaves the request queue (i.e., admitted into the task queue and started being served by at least one thread) and when it finally leaves the system (i.e., the first time when any $k$ of the corresponding tasks complete). In addition, we also consider the {\em task delays ($D_t$)}, which is the time it takes for a thread to serve a task {\bf assuming it is not terminated or canceled preemptively}. To clarify these definitions of delays, consider a request served with an $(n,k)$ MDS code, with $T_A$ its arrival time, $T_1\le T_2\le \cdots \le T_n$ the starting times of the corresponding $n$ tasks\footnote{We assume $T_i = \infty$ if the $i$-th task is never started.}. Then the queueing delay is $D_q = T_1 - T_A$. Suppose $D_{t,1},\cdots,D_{t,n}$ are the corresponding task delays, then the completion times of these task will be $X=\{T_1+D_{t,1},\cdots,T_n+D_{t,n}\}$ if none is canceled. So the request will leave the system at time $X_{(k)}$, which denotes the $k$-th smallest value in $X$, i.e., the time when $k$ tasks complete. Then the service delay of this request is $D_s = X_{(k)} - T_1$.

\section{Variable Chunk Sizing}
\label{sec:measurement}

\ifnewtext
In this section, we discuss implementation issues as well as pros and cons of two potential approaches, namely {\em Unique Key} and {\em Shared Key}, for supporting erasure-code-based access to files on the storage cloud with a variety of chunk sizes. Suppose the maximum desired redundancy ratio is $r$, then these approaches implement variable chunk sizing as follows:
\begin{itemize}
\item {\bf Unique Key:} For every choice of chunk size (or equivalently $k$), a separate batch of $rk$ coded chunks are created and each coded chunk is stored as an individual object with its unique key on the storage cloud. The access to different chunks is implemented through basic {\tt get, put} storage cloud APIs. 

\item {\bf Shared Key:} A coded file is first obtained by stacking together the coded strips obtained by applying a high-dimension $(N=rK,K)$ MDS code to the original file, as described in Section \ref{ssec:model:code} and illustrated in Fig.\ref{fig:partialRead}. 
For read, the coded file is stored on the cloud as one object. Access to chunks with variable size is realized by downloading segments in the coded file corresponding to batches of a corresponding number of strips, using a same key with more advanced ``partial read'' storage cloud APIs. Similarly, for write, the file is uploaded in parts using ``partial write'' APIs and then later merged into one object in the cloud.
\end{itemize}

\subsection{Implementation and Comparison of the two Approaches}
\label{ssec:measurement:partialRead}

\subsubsection{Storage cost} When the user request is to write a file, storage cost of Unique Key and Shared Key is not so different. However, to support variable chunk sizing for read requests, Shared Key is significantly more cost-efficient than Unique Key. With Shared Key, a single coded file stored on the cloud can be reused to support essentially an arbitrary number of different chunk sizes, as long as the strip size is small enough. On the other hand, it seems impossible to achieve similar reusing with the Unique Key approach where different chunks of the same file is treated as individual objects. So with Unique Key, every additional chunk size to be supported requires an extra storage cost $r\times$ file size. Such linear growth of storage cost easily makes it prohibitively expensive even to support a small number of chunk sizes.

\subsubsection{Diversity in delays} The success of \ourproposal and other 
proposals to use redundant requests (either with erasure coding or replication) for delay improvement relies on diversity in cloud storage access delays. In particular, \ourproposal, as well as \cite{fastcloud,Longbocodeingincloud,MDS-queue}, requires access delays for different chunks of {\bf the same file} to be weakly correlated.
  
With Unique Key, since different chunks are treated as individual objects, there is no inherent connection among them from the storage cloud system's perspective. So depending on the internal implementation of object placement policy of the storage cloud system, chunks of a file can be stored on the cloud in different storage units (disks or servers) on the same rack, or in different racks in the same data center, or even to different data centers at distant geographical locations. Hence it is quite likely that delays for accessing different chunks of the same file show very weak correlation. 

On the other hand, with Shared Key, since coded chunks are combined into one coded file and stored as one object in the cloud, it is very likely that the whole coded file, hence all coded chunks/strips, is stored in the same storage unit, unless the storage cloud system internally divides the coded file into pieces and distributes them to different units. Although many distributed storage systems do divide files into parts and store them separately, it is normally only for larger files. For example, the popular Hadoop distributed file system by default does not divide files smaller than 64MB. When different chunks are stored on the same storage unit, we can expect higher correlation in their access delays. It then is to be verified that the correlation between different chunks with the Shared Key approach is still weak enough for our coding solution to be beneficial.

\subsubsection{Universal support}
Unique Key is the approach adopted in our previous work \cite{fastcloud} to support erasure-code based file accessing with {\bf one predetermined} chunk size. A benefit of Unique Key is that it only requires basic {\tt get, put} APIs that all storage cloud system must provide. So it is readily supported by all storage cloud systems and can be implemented on top of any one.

On the other hand, Shared Key requires more advanced APIs that allow the proxy to download or upload only the targeted segment of an object. Such advanced APIs are not currently supported by all storage cloud systems. For example, to the best of our knowledge currently Microsoft's Azure Storage provides only methods for  ``partial read''\footnote{E.g. {\tt DownloadRangeToStream(target, offset, length)} downloads a segment of {\tt length} bytes starting from the {\tt offset}-th byte of the {\tt target} object (or ``blob'' in Azure's jargon).}
 but none for ``partial write''. On the contrary, Amazon S3 provides partial access for both read and write: the proxy can download a specific inclusive byte range within an object stored on S3 by calling {\tt getObject(request,destination)}\footnote{The byte range is set by calling {\tt request.setRange(start,end)}.}; and for uploading an {\tt uploadPart} method to upload segments of an object and an {\tt completeMultipartUpload} method to merge the uploaded segments are provided. We expect more service providers to introduce both partial read and write APIs in the near future.

\comment{

Nowadays, almost all popular storage cloud systems implement their own internal object replication mechanisms for very high availability and durability. These internal replication mechanisms usually try to place copies of an object to different storage units (disks or servers) in the same rack, different racks in the same data center, or even different data centers at distant geographical locations so that it is very unlikely that different copies will face the same network bottleneck and/or lost due to unexpected disaster. As a results,  

}

%
%\begin{table}[t]
%\centering
%\begin{tabular}{c|c}
%1123 & 123 \\
%123 & 123
%\end{tabular}
%\caption{Pros and Cons of Unique Key and Partial Access}
%\label{tab:pro&con}
%\end{table}

%
\comment{
In FAST CLOUD, the chunk size for each file is assumed to be predetermined and fixed, so it is plausible and cost-effective to treat each coded chunk as a individual object. However, this approach (referred as ``unique key'' hereafter) will easily be too expensive if more than one chunk sizes are to be supported because its storage cost grows linearly to the number of chunk sizes. For example, suppose we want to support downloading of a 3MB file using 3 chunk sizes of 3MB (no chunking), 1.5MB and 1MB ($k = 1, 2, 3$) and a maximum redundancy ratio of 2 ($n\le 2k$). One needs to create 2 objects of 3MB, 4 objects of 1.5MB and 6 objects of 1MB, resulting in a total storage cost of 18MB which is $6\times$ of the original file size instead of the desired redundancy ratio of $2$.
}

\fi

\subsection{Measurements on Amazon S3}
\label{ssec:measurement:S3}
%While the partial access approach appears to be an efficient way to support variable chunk sizing, it is important to also check its effectiveness: (1)  delays using such partial access functionality still demonstrate the two properties stated at the beginning of this section; and (2) it achieves delay improvement comparable to that from using the unique key approach. 
%
To understand the trade-off between Unique Key and Shared Key, we run measurements over Amazon EC2 and S3. EC2 instance served as the proxy in our system model. We instantiated an extra large EC2 instance with high I/O capability in the same availability region as the S3 bucket that stores our objects. We conducted experiments on different week days in May to July 2013 with various chunk sizes between 0.5MB to 3MB and up to $n=12$ coded chunks per file. For each value of $n$, we allow $L=n$ simultaneously active threads while the $i$-th thread being responsible for downloading the $i$-th coded chunk of each file. Each experiment lasted longer than 24 hours. We alternated between different settings to capture similar time of day characteristics across all settings. 
%For the same reasons, we also alternated between write and read jobs by first creating a batch of write jobs using distinct keys, then creating a batch of read jobs for these distinct keys once all the writes are completed successfully. 

The experiments are conducted within all 8  availability regions in Amazon S3. 
Except for the ``US Standard'' availability region, all other 7 regions demonstrate similar performance statistics that are consistent over different times and days. We conjecture the different and inconsistent behavior of ``US Standard'' might be due to the fact that it targets a slightly different usage pattern and it may employ a different implementation for that reason\footnote{See \url{http://docs.aws.amazon.com/general/latest/gr/rande.html#s3_region}}. We will exclude ``US Standard'' from subsequent discussions. 
Due to lack of space, we only show a limited subset of findings for availability region ``North California'' that are representative for regions other than ``US Standard'':

\begin{figure}[!t]
\centering
\includegraphics[width= \onewidth]{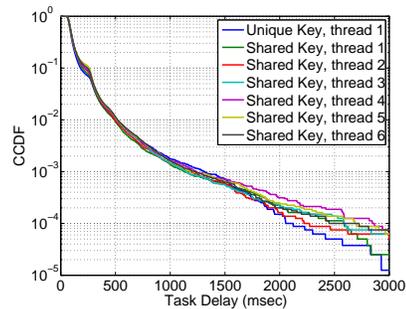}
\vspace{\shrinkbeforecaption}
\caption{CCDF of individual threads with 1MB chunks and $n=6$}
\label{fig:ccdf:thread}
\vspace{\shrinkaftercaption}
\end{figure}

(1)
\label{obs:identicalDistribution} 
In both Unique Key and Shared Key, the task delay distribution observed by different threads are almost identical. The two approaches are indistinguishable  even beyond 99.9th percentile. Fig.\ref{fig:ccdf:thread} show the complementary cumulative distribution function (CCDF) of task delays observed by individual threads for 1MB chunks and $n=6$. Both approaches demonstrate large delay spread in all regions.

\begin{figure}[!t]
\centering
	\subfigure[Unique key]{
		\label{fig:ccdf:uniqueKey}
		\includegraphics[width=\twowidth ]{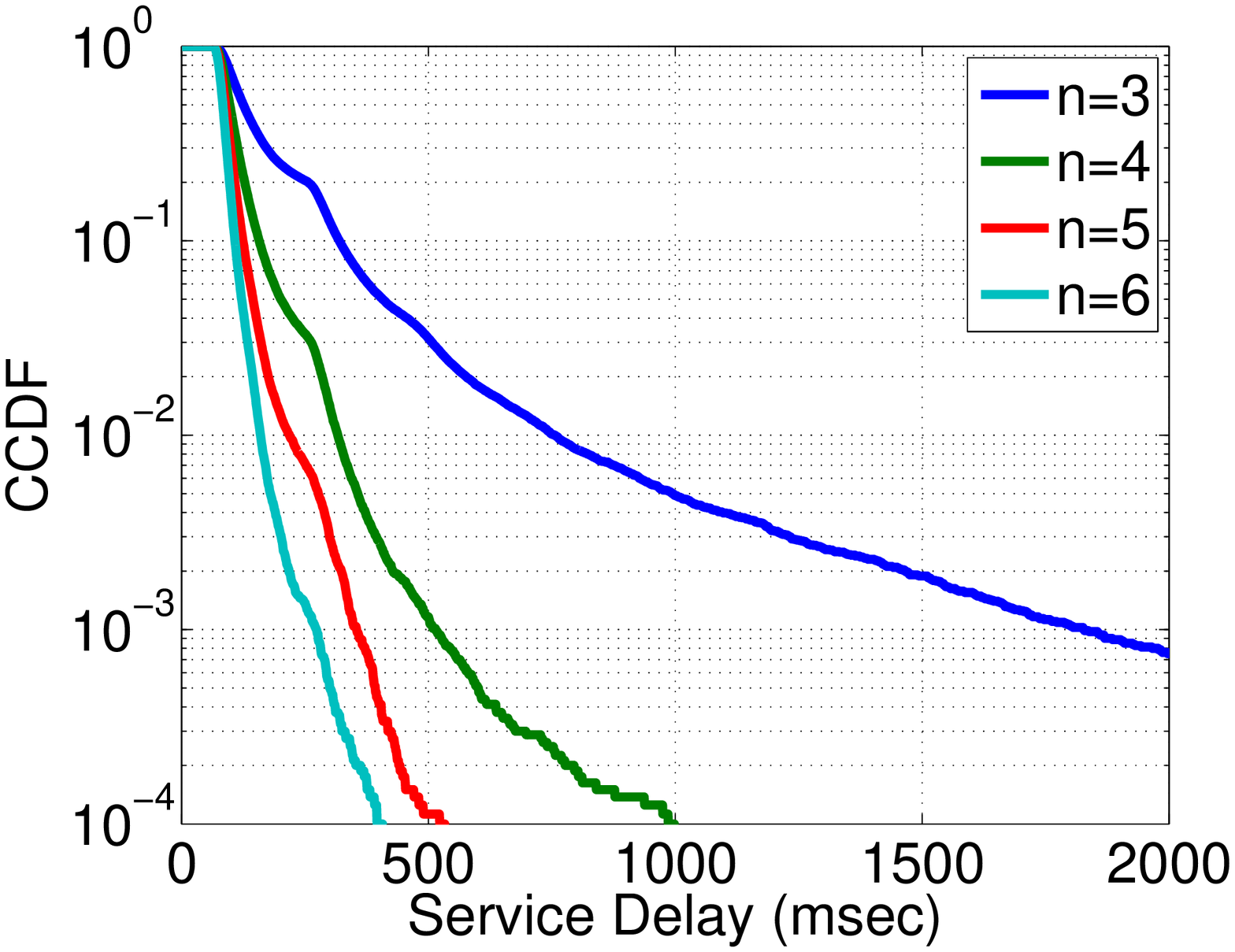}
	}%
	\subfigure[Partial read]{
		\label{fig:ccdf:partialRead}
		\includegraphics[width=\twowidth ]{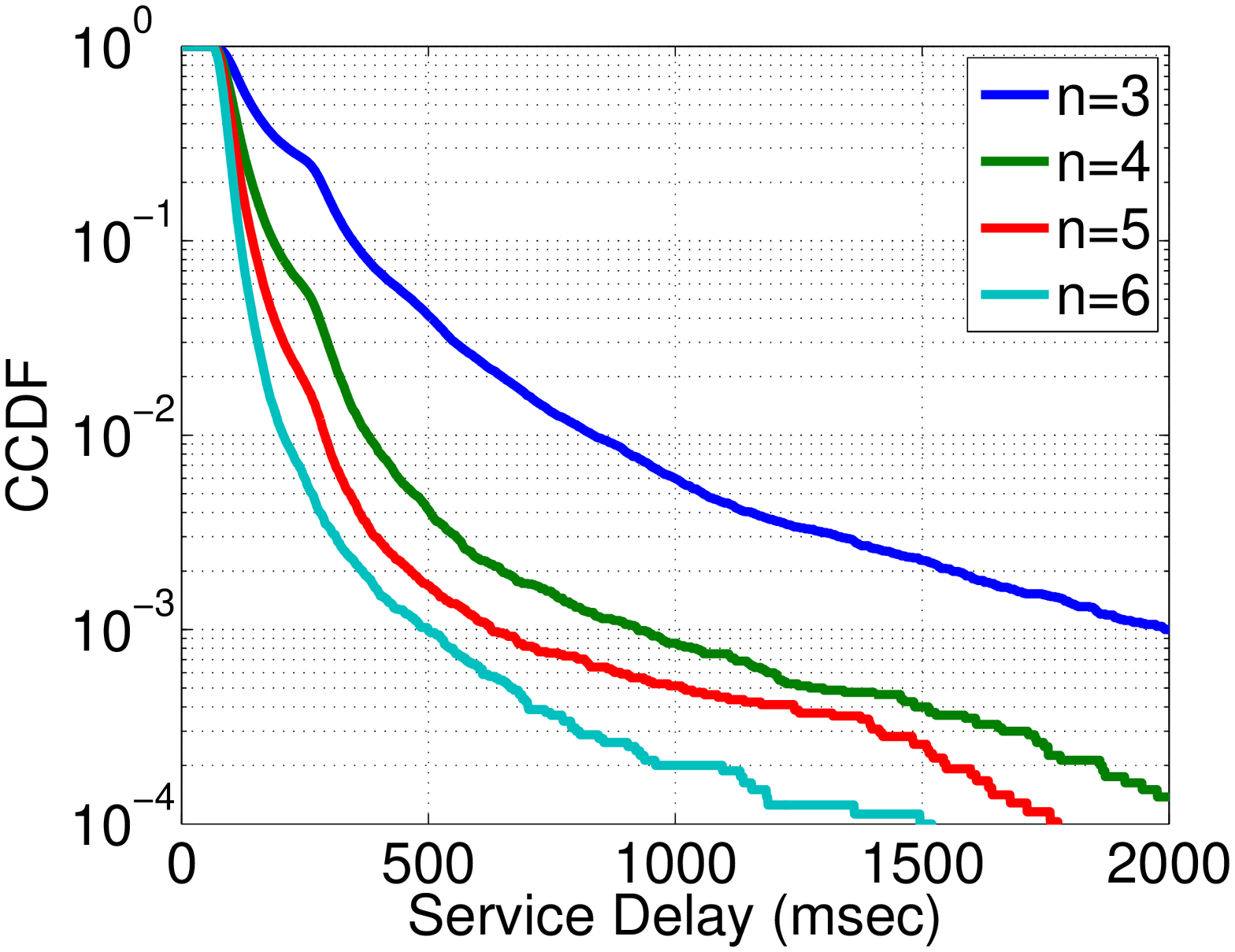}
	}
\vspace{\shrinkbeforecaption}
\caption{CCDF of service delay for reading 3MB files with 1MB chunks}
\label{fig:CCDF:FEC}
\vspace{\shrinkaftercaption}
\end{figure}

(2)
\label{obs:weakCorrelation}
Task delays for different threads in Unique Key show close to zero correlation, while they demonstrate slightly higher correlation in Shared Key, as it is  expected.
At all different settings, the cross correlation coefficient between different threads stays below 0.05 in Unique Key and ranges from 0.11 to 0.17 in Shared Key. 
Both approaches achieve significant service delay improvements. 
Fig.\ref{fig:CCDF:FEC} plots the CCDF of service delays for downloading 3MB files with 1MB chunks ($k=3$) with $n= 3 \sim 6$, assuming all $n$ tasks in a batch start at the same time. In this setting, both approaches reduce 99th percentile delays by roughly 50\%, 65\% and 80\% by downloading 1, 2 and 3 extra coded chunks. Although Shared Key demonstrates up to 3 times higher cross correlation coefficient, there is no meaningful statistical distinction in service delay between the two approaches until beyond 99th percentile.
All availability regions experience different degrees of degradation at high percentiles with Shared Key due to the higher correlation. 
Significant degradation emerges from around 99.9th percentile and beyond in all regions except for ``Sao Paulo'', in which degradation appears around 99th percentile.

(3)
\label{obs:lowerBound} 
Task delays are always lower bounded by some constant $\Delta\ge 0$ that grows roughly linearly as chunk size increases. This constant part of delay cannot be reduced by using more threads: see the flat segment at the beginning of the CCDF curves in Fig.\ref{fig:ccdf:thread} and Fig.\ref{fig:CCDF:FEC}. Since this constant portion of task delays is unavoidable, it leads to the negative effect of using larger $n$ since there is a minimum cost of system resource of $n\Delta $ (time$\times$thread) that grows linearly in $n$. This cost leads to a reduced capacity region for using more redundant tasks, as illustrated in the example of Fig.\ref{fig:fixedDelays}. 
We observe that the two approaches deliver almost identical total delays (queueing + service) for all arrival rates, in spite of the degraded service delay with Shared Key at very high percentile. So we only plot the results with Shared Key in Fig.\ref{fig:fixedDelays}.

\begin{figure}[!t]
\centering
	\subfigure[Mean]{
		\label{fig:meanTaskDelay}
		\includegraphics[width= \twowidth ]{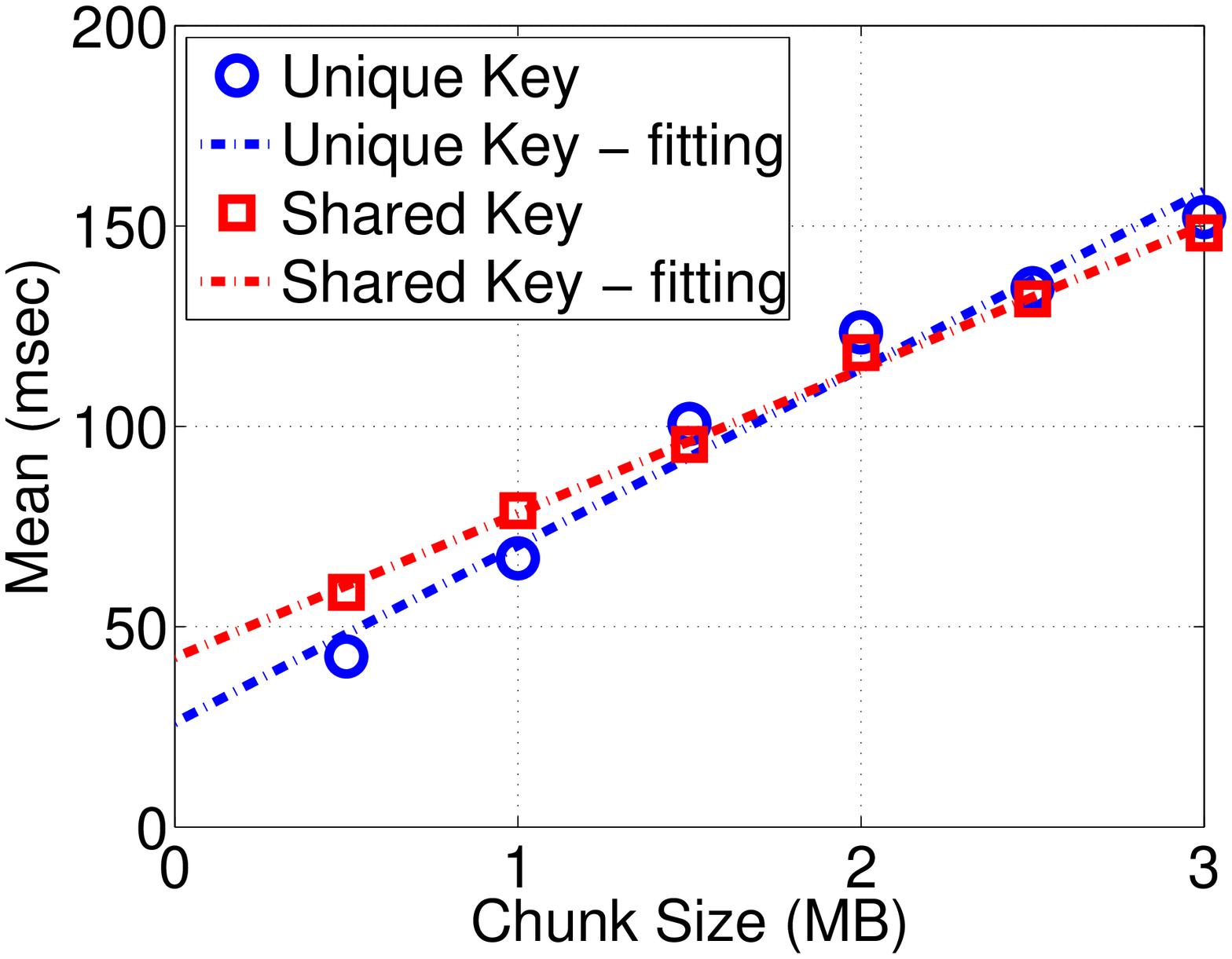}
	}%
	\subfigure[Standard Deviation]{
		\label{fig:stdTaskDelay}
		\includegraphics[width= \twowidth ]{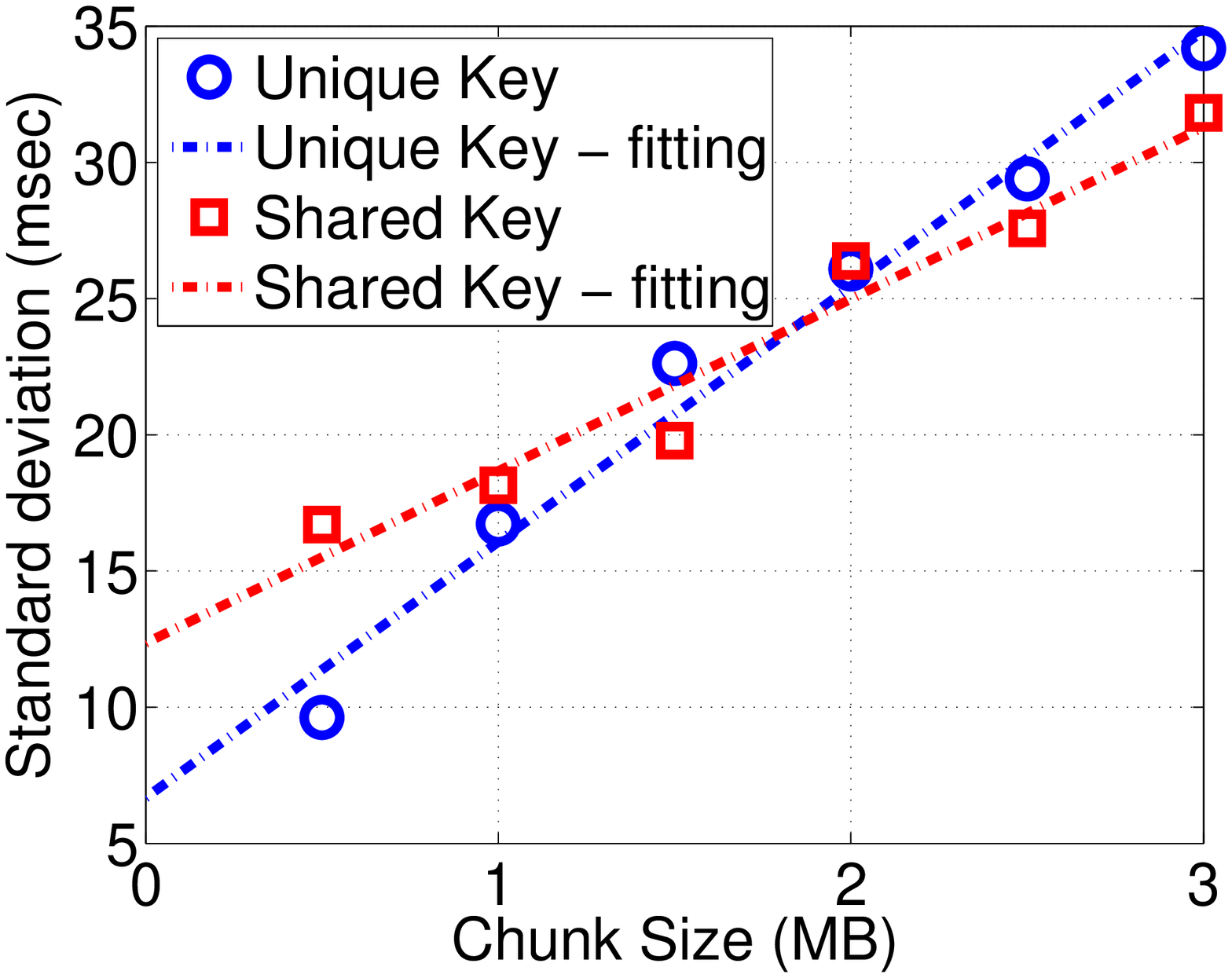}
	}
\vspace{\shrinkbeforecaption}
\caption{Delay statistics vs. chunk size}
\label{fig:delay-chunkSize}
\vspace{\shrinkaftercaption}
\end{figure}

(4)
\label{obs:linearGroth}
 The mean and standard deviation of task delays grows roughly linearly as chunk size increases. Fig.\ref{fig:delay-chunkSize} plots the measured mean and standard deviation of task delays in both approaches at different chunk sizes. Also plotted in the figures are least squares fitted lines for the measurement results. Notice that the extrapolations at chunk size = 0MB are all greater than zero. We believe this observation reflects the costs of non-I/O-related operations in the storage cloud that do not scale proportionally to object size, for example the cost to locate the requested object. We also believe such costs contribute partially to the minimum task delay constant $\Delta$.

\subsection{Model of Task Delays}
\label{ssec:model:delay}
Based on the aforementioned observations, we decide to use the Shared Key approach in \ourproposal since its outstanding storage efficiency overweights the minimum degradation in delay. For the analysis present in the next section,
we model the task delays as independently distributed random variables whose mean and standard deviation grow linearly as chunk size $B$ increases. More specifically, we assume the task delay $D_t$ for chunk size $B$ following distribution in the form of 
\begin{equation}
D_t(B) \sim \Delta(B) + exp(\mu(B)),
\end{equation}
where 
$
\Delta(B) = \fixedDelta + \linearDelta B
$
captures the lower bound of task delays as in observation (3), and $exp(\mu(B))$ represents a exponential random variable that models the tail of the CCDF. The mean and standard deviation of the exponential tail both equal to  
$
\frac{1}{\mu_i(B)} = \fixedExp + \linearExp B.
$
With this model, constants
$\fixedDelta$ and $\fixedExp$ together capture the non-zero extrapolations of the mean and standard deviation of task delays at chunk size 0, and similarly, constants $\linearDelta$ and $\linearExp$ together capture the rate at which the mean and standard deviation grow as chunk size increases, as in observation (4).

\comment{
We assume there are $m\ge 1$ classes of requests. Requests of each class have identical file size  and all are divided into chunks of identical size. Under this assumption, service times of all chunks of the same class follow the same distribution and each class $i$ can be characterized by a three-tuple $(k_i,\Delta_i,\mu_i)$, where $\Delta_i$ and $\mu_i$ specifies the delay distribution of class-$i$ chunks.
Throughout this paper, we assume $k_i$'s (and accordingly chunk sizes) are determined a priori and $(\Delta_i,\mu_i)$ are given. Our focus will be on the adaptation/choice of $n_i$'s.  
}

\section{Design of \ourproposal}
\label{sec:proposed-algorithm}

For the analysis in this section, we group requests into classes according to the tuple {\tt (type, size)}. Here {\tt type} can be read or write, and can potentially be other type of operations supported by the cloud storage. Each type of operation has its own set of delay parameters $\{\fixedDelta, \linearDelta, \fixedExp, \linearExp\}$. Subscripts will be used to indicate variables associated with each class.  

We first introduce approximations for the expected queueing and service delays, assuming the FEC code used to serve requests of each class is predetermined and fixed (Section \ref{ssec:ana:static})
Then we formulate an optimization problem whose objective is to minimize the expected total delay over all such static strategies with fixed FEC codes. We show that solutions to the non-convex optimization problem exhibit a nice property (Section \ref{ssec:ana:optStatic}):

\mybox{
The optimal values of $n_i$, $k_i$ and $r_i$ can all be expressed as functions of solely determined by $Q$ -- the expected length of the request queue:
$$n_i = N_i(Q),~~k_i = K_i(Q) \text{~~and~~}  r_i = R_i(Q).$$
$N_i$, $K_i$ and $R_i$ are all strictly decreasing functions of $Q$.
}
%\begin{enumerate}
%\item The optimal choice of FEC codes can be expressed as a function solely determined by the corresponding expected queue length $Q$ of the request queue (ignoring type (read/write/etc.) and size of requests), without explicitly knowing the arrival rate and composition of different requests. 
%
%\item Both code dimension $k$ and redundancy ratio $r$ are strictly decreasing functions of the expected queue length $Q$. 
%\end{enumerate}

\noindent This finding is then used as the guideline in the design of our backlog-driven adaptive strategy \ourproposal (Section \ref{ssec:ana:adaptive}).

\subsection{Approximated Analysis of Static Strategies}
\label{ssec:ana:static}
Denote $J_i$ as the file size of class $i$.
Consider a request of class $i$ served with an $(n_i,k_i)$ MDS code, i.e., $B_i = J_i/k_i$. First suppose {\em all $n_i$ tasks start at the same time}, i.e., $T_1 = T_{n_i}$. In this case, given our model for task delays, it is trivial to show that the expected service delay equals to 
\begin{eqnarray}
D_{s,i} &=& \Delta_i(J_i/k_i)
				+ \frac{1}{\mu_i(J_i/k_i)}
				  \sum_{j=0}^{k_i-1}\frac{1}{n_i - j}
\nonumber\\
&\approxeq& \Delta_i(J_i/k_i) 
				+ \frac{1}{\mu_i(J_i/k_i)}\ln\frac{n_i}{n_i-k_i} 
\nonumber\\
&=& \fixedDelta_i + \frac{\linearDelta_i J_i}{k_i} + 
\left(\fixedExp_i + \frac{\linearExp_i J_i}{k_i}\right)\ln\frac{r_i}{r_i-1}.
\end{eqnarray}
Also define the system usage (or simply cost) of a request as the sum of the amount of time each of its tasks being served by a thread\footnote{The time a task $j$ being served is $D_{t,j}$ if it completes successfully, $X_{(k)}-T_j$ if it starts but is terminated preemptively, and 0 if it is canceled while waiting in the task queue.}. When all tasks start at the same time, its expected system usage is (see Section IV of \cite{fastcloud} for detailed derivation)
\begin{eqnarray}
U_i &=& n_i\Delta_i(J_i/k_i) + \frac{k_i}{\mu_i(J_i/k_i)}\nonumber\\
 &=& \fixedDelta_i k_i r_i + \linearDelta_i J_i r_i 
+ \fixedExp_i k_i + \linearExp_i J_i.
\end{eqnarray}

Suppose class $i$ contributes to $p_i$ fraction of the total arrivals, then the average cost per request is $\aveUsage = \sum_i p_i U_i$. With $L$ simultaneously active threads, requests depart the system at rate $L/\aveUsage$ (request/unit time). In light of this observation, we approximate the request queue with an $M/M/1$ queue with service rate $L/\aveUsage$. So the queueing delay in the original system at total arrival rate $\lambda$ is approximated by 
\begin{eqnarray}
D_q = \frac{1}{L/\aveUsage - \lambda} - \frac{1}{L/\aveUsage}
= \frac{\lambda \aveUsage^2}{L(L-\lambda \aveUsage)},
\end{eqnarray}
and the expected  length of the request queue is approximately 
\begin{equation}
Q = \lambda D_q= \frac{(\lambda \aveUsage)^2}{L(L-\lambda \aveUsage)} = \frac{\normArrival^2}{L(L-\normArrival)}.
\label{eq:Q}
\end{equation}
Here $\normArrival = \lambda \aveUsage = \lambda  \sum_i p_i U_i
= \lambda \sum_i p_i (\fixedDelta_i k_i r_i + \linearDelta_i J_i r_i 
+ \fixedExp_i k_i + \linearExp_i J_i)$.

We acknowledge that the above approximation is quite coarse, especially because tasks of the same batch do not start at the same time in general.  However, remember the main objective of this paper is to develop a practical solution that can achieve the optimal delay-throughput trade-off. According to the simulation results, this approximation is sufficiently good for the purpose of this paper.

\subsection{Optimal Static Strategy}
\label{ssec:ana:optStatic}
Given total arrival rate $\lambda$ and composition of requests $\{p_i\}$, we want to find the best choice of FEC code for each class such that the total delay is minimized. Relaxing the requirement for $n_i$ and $k_i$ being integers, this is formulated as the following minimization problem\footnote{Notice that all classes share the same queueing delay. Also, we require $k_i>0$ instead of $k_i\ge 1$ for a technicality to simplify the proof of the uniqueness of the optimal solution. We require $r_i\ge 1$ since $n_i\ge k_i$. $\normArrival < L$ is imposed for queue stability.}:
\begin{align*}
\min_{\{k_i,r_i\}} &~~~ D_q + \sum_i p_i D_{s,i} 
\label{eq:optimization}
\tag{*}
\\
s.t. &~~~ k_i>0,~~ r_i \ge 1,~~ \normArrival < L. 
\end{align*}

Notice that this is a non-convex optimization problem because the feasible region is not a convex set, due to the $k_i r_i$ terms in $\normArrival$. In general, non-convex optimization problems are difficult to solve. Fortunately, we are able to prove the following theorem according to which this non-convex optimization problem can be solved numerically efficiently.

\begin{theorem}
For any given $\lambda$ and $\{p_i\}$, the non-convex optimization problem \eqref{eq:optimization} has a unique optimal solution, which satisfies the following for all $i$:
\begin{align}
\frac{k_i(\fixedExp_i k_i  + \linearExp J_i)}{\fixedDelta_i k_i + \linearDelta_i J_i} 
&= \frac{J_i r_i(r_i-1)}{\fixedDelta_i r_i + \fixedExp_i}\left(\linearDelta_i + \linearExp_i\ln\frac{r_i}{r_i-1}\right),
\label{eq:opt:code}
\\
\left(\frac{L}{L-\normArrival}\right)^2-1
&= \frac{2L (\fixedExp_i k_i + \linearExp_i J_i)}{k_i r_i (r_i-1) (\fixedDelta_i k_i + \linearDelta_i J_i)}.
\label{eq:opt:normArrival}
\end{align}
\end{theorem}
\begin{IEEEproof} See Appendix.
\end{IEEEproof}

Observing that Eq.\ref{eq:opt:code} contains only delay parameters and file size of class $i$, so it should be always satisfied no matter what arrival rate $\lambda$ and request composition $\{p_i\}$ are. Solving Eq.\ref{eq:opt:code} alone gives a set of pairs $(k_i,r_i)$ that are the optimal choice of code for class $i$ {\bf for some} $\lambda$ and $\{p_i\}$. Then solving Eq.\ref{eq:opt:normArrival} within this set we obtain the optimal $k_i$ and $r_i$ as a function of $\normArrival$ {\bf for all} combinations of $\lambda$ and $\{p_i\}$ such that $\normArrival = \lambda \sum_i p_i U_i$. Observing from Eq.\ref{eq:Q} that $\normArrival = L\left(\sqrt{Q^2+4Q}-Q\right)/2$, and with some simple calculus, we conclude that 
\begin{corollary}
\label{corr:func_of_Q}
The optimal values of $n_i$, $k_i$ and $r_i$ can all be expressed as strictly decreasing functions of $Q$:
\begin{equation}
n_i = N_i(Q),~~k_i = K_i(Q) \text{~~and~~}  r_i = R_i(Q).
\label{eq:func_of_Q}
\end{equation}
\end{corollary}

\subsection{Adaptive Strategy \ourproposal}
\label{ssec:ana:adaptive}
The finding of Corollary \ref{corr:func_of_Q} conforms to our intuition: 
\begin{itemize}
\item At light workload (small $\lambda$), there should be little backlog in the request queue (small $Q$) and the service delay dominates the total delay. In this case, the system is not operating in the throughput-limited regime, so it is beneficial to increase the level of chunking and redundancy. 

\item At heavy workload (larger $\lambda$), there will be a large backlog in the request queue (large $Q$) and the queueing delay dominates the total delay. In this case, the system operates in the throughput-limited regime, So it is better to reduce the level of chunking and redundancy to support higher throughput.
\end{itemize}

More importantly, it suggests that it is sufficient to choose the FEC code solely based on the length of the request queue. The basic idea of \ourproposal is to choose $n_i = N_i(q)$ and $k_i= K_i(q)$ for a request of class $i$, where $q$ is the queue length upon the arrival of the request. When this is done to all requests arrive into the system, it can be expected the average code lengths (dimensions) and expected queue length $Q$ satisfy Eq.\ref{eq:func_of_Q}, hence optimal delay is achieved.  In \ourproposal, this is implemented with a threshold based algorithm, which can be performed very efficiently. 
For each class $i$, we first compute the expected queue length if $n_i = 1,...,n_i^{max}$ is the optimal code length by
\begin{equation}
Q^{N}_{i,n_i} = N_i^{-1}(n_i).
\end{equation}
Here $n_i^{max}$ is the maximum number of tasks allowed for a class $i$ request. Since $N_i$ is a strictly decreasing function, its inverse $N_i^{-1}$ is a well-defined strictly decreasing function. As a result, we have
$ Q^{N}_{i,1} > Q^{N}_{i,2} > \cdots > Q^{N}_{i,n_i^{max}} >0.$
Remember our goal is to use code length $n$ if queue length $q$ is around $Q^{N}_{i,n}$, so we want a set of thresholds $\{\nthreshold_{i,n}\}$ such that
\begin{align*} 
\nthreshold_{i,1}> Q^{N}_{i,1} &>\nthreshold_{i,2} > Q^{N}_{i,2} > \cdots \\
\cdots &> \nthreshold_{i,n_i^{max}} > Q^{N}_{i,n_i^{max}} > \nthreshold_{i,n_i^{max}+1}=0,
\end{align*}
and will use $n$ such that $q\in[\nthreshold_{i,n+1},\nthreshold_{i,n})$.
In our current implementation of \ourproposal, we use
$ \nthreshold_{i,n} = \left(Q^{N}_{i,n} + Q^{N}_{i,n-1}\right)/2$
for  $n = 2,\cdots, n_i^{max}$ and $\nthreshold_{i,1} = \infty$.
A set of thresholds $\{\kthreshold_{i,k_i^{max}}\}$ for adaptation of $k_i$ is found in a similar fashion.
The adaptation algorithm of \ourproposal is summarized in pseudo-codes below:

~

\hrule
\vspace{2pt}
\noindent \textit{\textbf{\ourproposal (Throughput Optimal FEC Cloud)}}
\hrule
\vspace{2pt}

\noindent {\bf Initialization: $\overline{q} = 0$}

\noindent {\tt request} arrives
\begin{algorithmic} [1]
\STATE $q \leftarrow $ queue length upon arrival of {\tt request}
\STATE $i\leftarrow$ class that {\tt request} belongs to
\STATE $\overline{q} \leftarrow \alpha q + (1-\alpha)\overline{q}$
\STATE Find $k\le k_i^{max}$ such that $\overline{q}\in [\nthreshold_{i,k+1},\nthreshold_{i,k})$ 
\STATE Find $n\le n_i^{max}$ such that $\overline{q}\in [\nthreshold_{i,n+1},\nthreshold_{i,n})$ 
\STATE $n \leftarrow \min(r_i^{max}k, n)$ 
\label{step:double-check-n}
\STATE Serve {\tt request} with an $(n,k)$ code when it becomes HoL.
\end{algorithmic}
\hrule

~

Note that in Step \ref{step:double-check-n} we reduce $n$ to $r_i^{max}k$ if the redundancy ratio of the code chosen in the previous steps is higher than $r_i^{max}$ -- the maximum allowed redundancy ratio for class $i$. 
Also, instead of comparing $q$ directly with the thresholds, we compare an exponential moving average 
$\overline{q} = \alpha q + (1-\alpha)\overline{q}$,
with a memory factor $0< \alpha \le 1$, against the thresholds to determine $n$ and $k$. The moving average is used to mitigate the transient variation in queue length so that $n$ and $k$ will not change too frequently. It is obvious that we only need to set $\alpha=1$ in order to use instantaneous queue length $q$ for the adaptation since in this case $\overline{q}=q$.

\begin{figure*}[ht]
\centering
	\subfigure[Mean Delay]{
		\label{fig:read:ave}
		\includegraphics[width=\fourwidth]{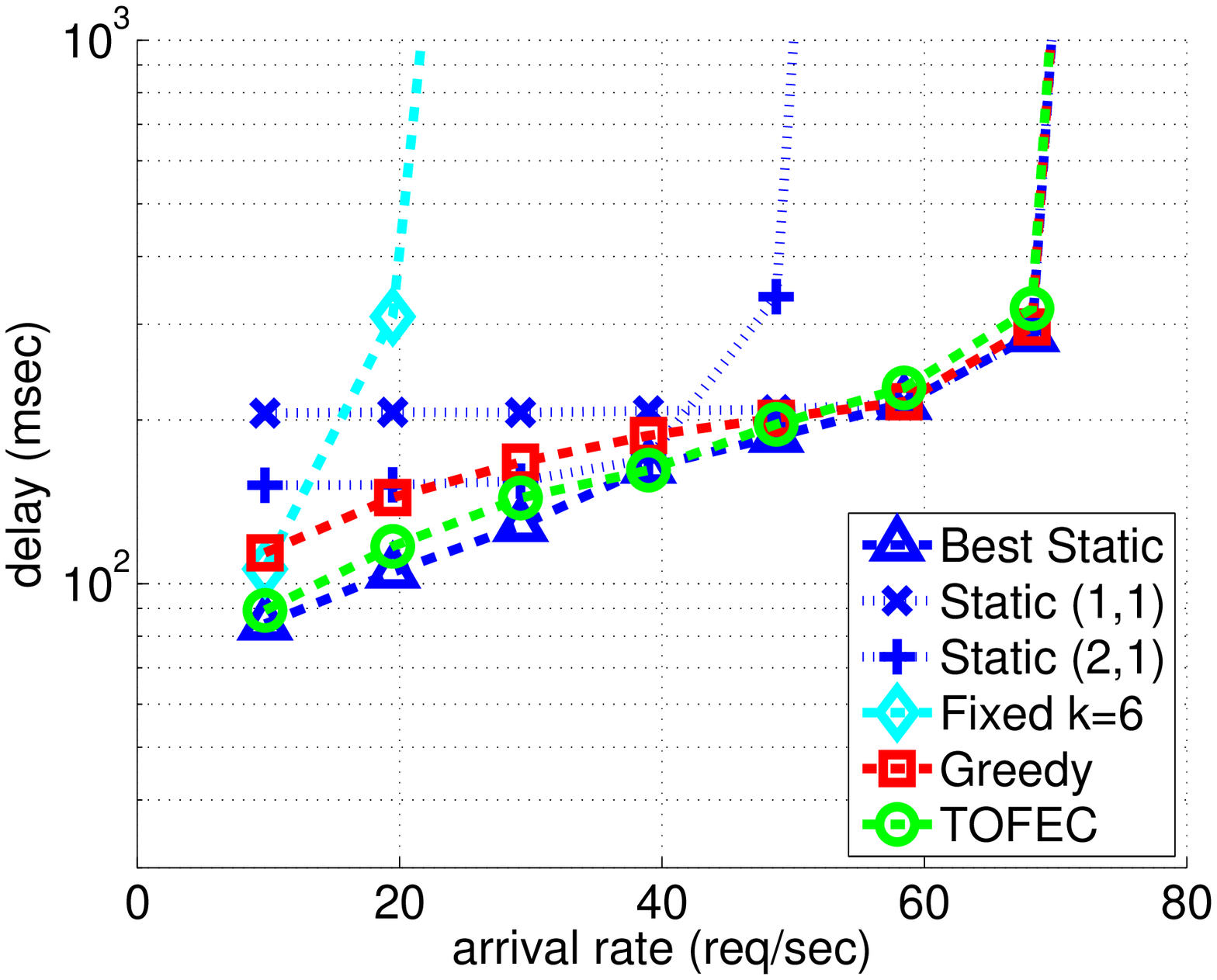}
	}%
~
	\subfigure[Median Delay]{
		\label{fig:read:med}
		\includegraphics[width=\fourwidth]{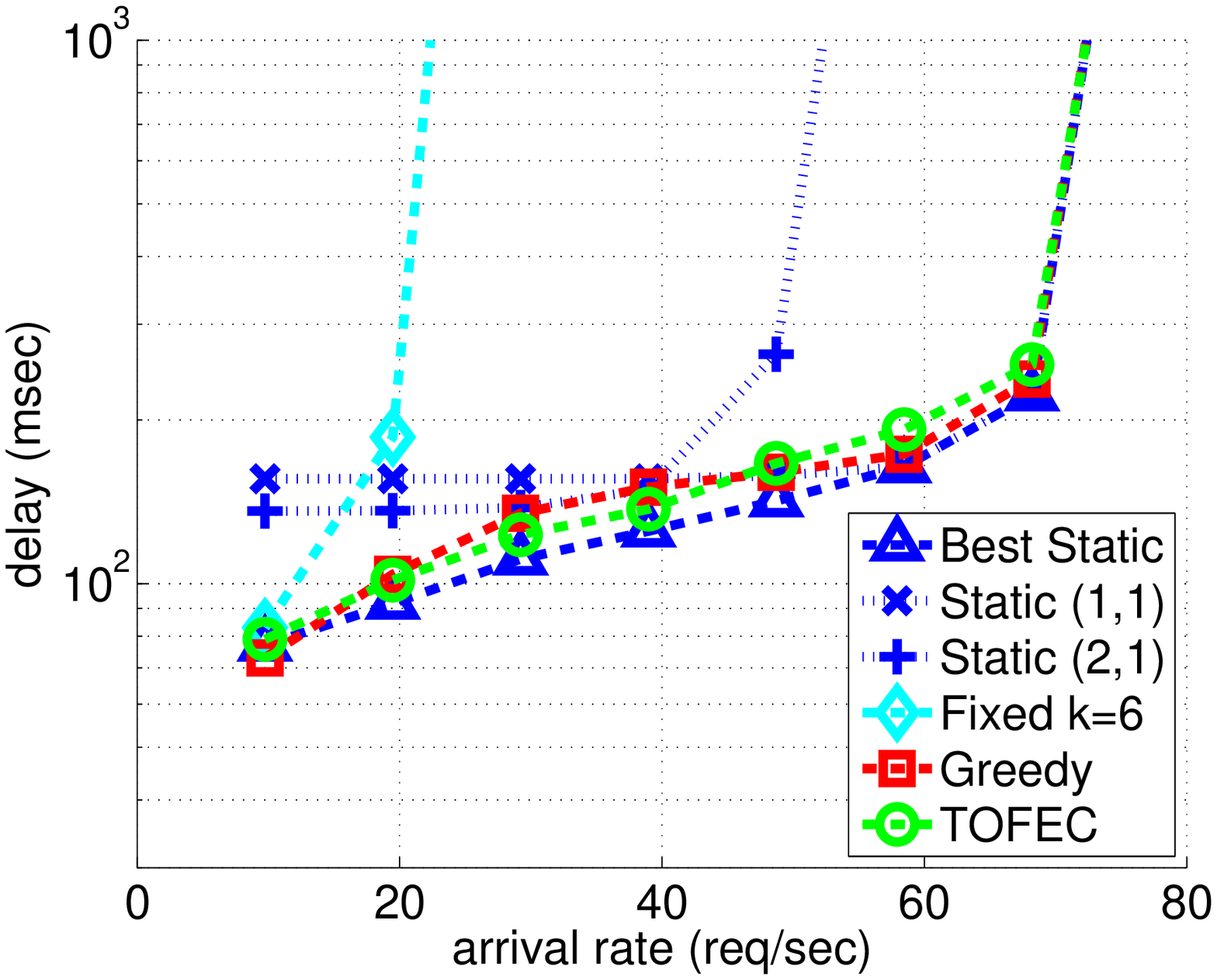}
	}
~
	\subfigure[90th Percentile Delay]{
		\label{fig:read:9}
		\includegraphics[width=\fourwidth]{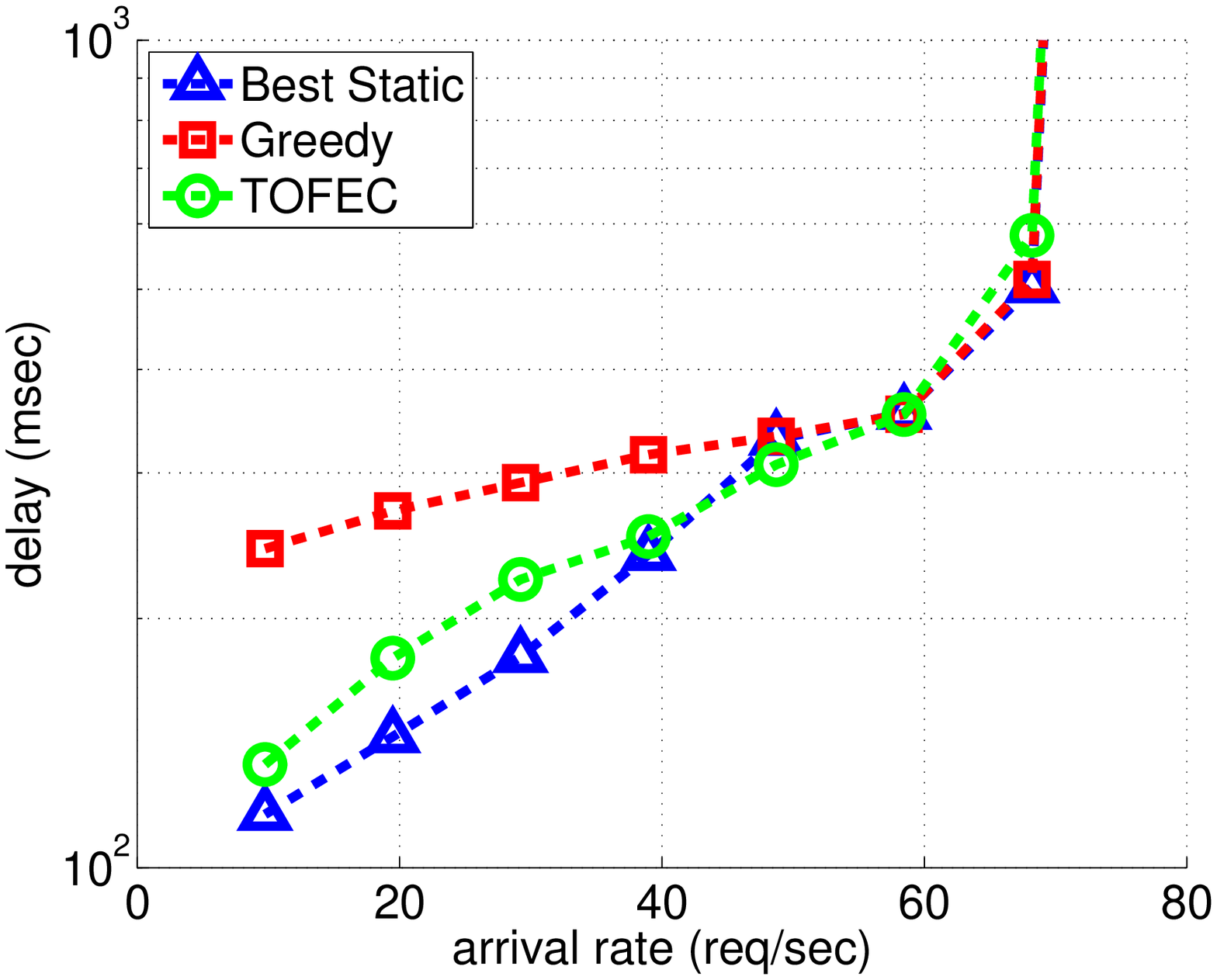}
	}
~
	\subfigure[99th Percentile Delay]{
		\label{fig:read:99}
		\includegraphics[width=\fourwidth]{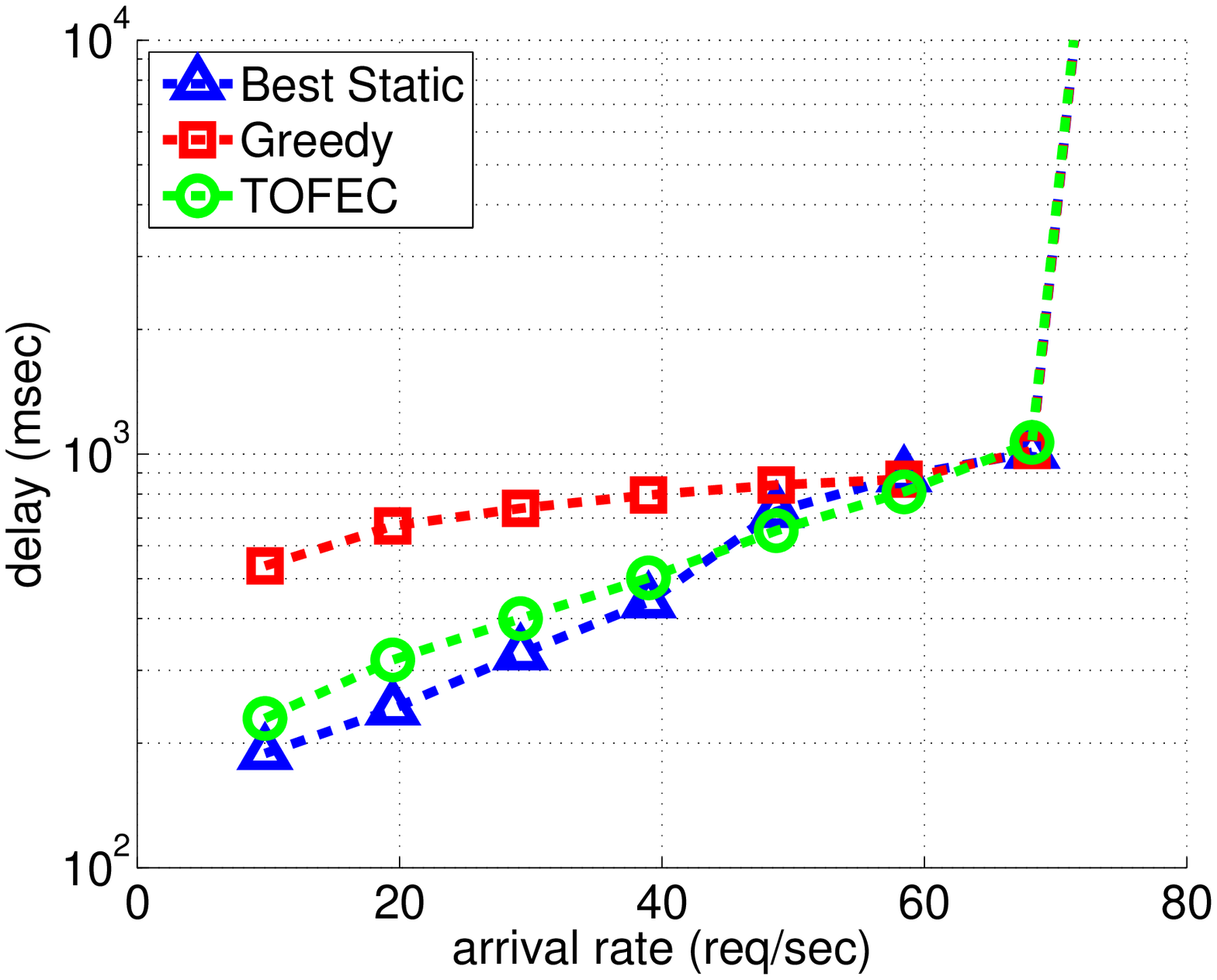}
	}
%~
%	\subfigure[Standard Deviation]{
%		\label{fig:read:std}
%		\includegraphics[width=\fourwidth]{read_std}
%	}
\vspace{\shrinkbeforecaption}
\caption{Delay performance in read only scenario}
\label{fig:read}
\vspace{\shrinkaftercaption}
\end{figure*}

\section{Evaluation}
\label{sec:evaluation}

We now demonstrate the benefits of \ourproposal 's adaptation mechanism. 
We evaluate \ourproposal 's adaptation strategy and show that is outperforms static strategies with both constant and changing workloads, as well as a simple greedy heuristic that will be introduced later.
%In Section \ref{ssec:eva:setup}, we describe the setup for our simulations. In Section \ref{ssec:eva:singleclass} -- \ref{ssec:eve:multiclass} 

\subsection{Simulation Setup}
\label{ssec:eva:setup}

We conducted trace-driven simulations for performance evaluation for both single-class and multi-class scenarios with both read and write requests of different file sizes. Due to lack of space, we only show results for the scenario with one class {\tt (read,3MB)}. But we must emphasize that it is representative enough so that the findings to be discussed in this section are valid for other settings (different file sizes, write requests, and multiple classes). We assume the system supports up to $L=16$ simultaneously active threads. We set the maximum code dimension and redundancy ratio to be $k^{max} = 6$ and $r^{max} = 2$, because we observe negligible gain in service delay beyond this chunking and redundancy level from our measurements. We use traces collected in May and June in availability region ``North California''. In order to compute the threshold for \ourproposal, we need estimations of the delay parameters $\{\fixedDelta,\linearDelta,\fixedExp,\linearExp\}$. For this, we first filter out the worst 10\% task delays in the traces, then we compute the delay parameters from the least squares linear approximation for the mean and standard deviation of the remaining task delays. We use memory factor $\alpha = 0.99$ in \ourproposal.

In addition to the static strategies, we develop a simple {\em Greedy} heuristic strategy for the purpose of comparison. Unlike the adaptive strategy in \ourproposal, Greedy does not require prior-knowledge of the distribution of task delays, yet it achieves competitive mean delay performance. In Greedy, the code to be used to serve request in class $i$ is determined by the number of idle threads upon its arrival: suppose there are $l$ idle threads, then
$
k_i = 
\begin{cases}
1, & \mbox{if } l=0\\
\min(k_i^{max},l), & \mbox{otherwise}
\end{cases},
$
and similarly
$
n_i = 
\begin{cases}
1, & \mbox{if } l=0\\
\min(r_i^{max}k_i,l), & \mbox{otherwise}
\end{cases}.
$
The idea of Greedy is to first maximize the level of chunking with the idle threads available, then increase the redundancy ratio as long as there are idle threads remain.

%\subsection{Single-Class Scenario}
%\label{ssec:eva:singleclass}
%We show the results in scenarios with only one class of requests (read-only or write-only). 

\subsection{Throughput-Delay Trade-Off}
Fig.\ref{fig:read} shows the mean, median, 90th percentile and 99th percentile delays of  \ourproposal and Greedy with Poisson arrivals at different arrival rate $\lambda$. We also run simulations with static strategies for all possible combinations of $(n,k)$ at every arrival rate. We brute-force find the best mean, median, 90th and 99th percentile delays achieved with static strategies and use them as the baseline. 
Also plot in Fig.\ref{fig:read:ave} and Fig.\ref{fig:read:med} are the mean and median delay performance of the basic static strategy with no chunking and no replication, i.e., $(1,1)$ code; the simple replication static strategy with a $(2,1)$ code; and the backlog-based adaptive strategy from \cite{fastcloud} with fixed code dimension $k=6$ and $n\le 12$. 

As we can see, \ourproposal and Greedy have almost identical performance in terms of mean and median delays. Both \ourproposal and Greedy achieve (almost) optimal mean and median delays for all arrival rates.
At light workload, \ourproposal delivers about $2.5\times$ improvement in mean delay when compared with the basic static strategy, and about $2\times$ when compared with simple replication (from 205ms and 151ms to 84ms). It also reduces the median delay by about $2\times$ from that of basic and simple replication (from 156ms and 138ms to 74ms). Meanwhile Greedy achieve about $2\times$ improvement in both mean (89ms) and median delays (79ms) over basic. 

With heavier workload, both \ourproposal and Greedy successfully adapt their codes to keep track with the best static strategies, in terms of mean and median delays. It is clear from the figures that both \ourproposal and Greedy achieve our primary goal of retaining full system capacity, as supported by basic static strategy. On the contrary, although simple replication has slightly better mean and median delays than basic under light workload, it fails to support arrival rates beyond 70\% of the capacity of basic.
Meanwhile, the adaptive strategy from \cite{fastcloud} with fixed code dimension $k=6$ can only support less than 30\% of the original capacity region, although it achieves the best delay at very light workload.

\comment{
\begin{figure}[t]
	\subfigure[Average $k$]{
		\label{fig:read:codeDim}
		\includegraphics[width = \twowidth]{read_codeDim}
	}%
	\subfigure[Average $n$]{
		\label{fig:read:codeLength}
		\includegraphics[width = \twowidth]{read_codeLength}
	}
\vspace{\shrinkbeforecaption}
\caption{Delay performance in read only scenario}
\label{fig:read:code}
\vspace{\shrinkaftercaption}
\end{figure}
}

%\subsubsection{High Percentile Delays}
While the two adaptive strategies have similar performance in mean and median, \ourproposal outperforms Greedy significantly at high percentiles. As Fig.\ref{fig:read:9} and Fig.\ref{fig:read:99} demonstrate, \ourproposal matches with the best static strategies at 90th and 99th percentile delays throughout the whole capacity region. On the other hand, Greedy fails to keep track of the best static performance at lower arrival rates. At light workload, \ourproposal 's is over $2\times$ and $2.5\times$ better than Greedy at 90th and 99th percentiles. Less interesting is the case with heavy workload and the system is throughput-limited. Hence both strategies converge to the basic static strategy using mostly $(1,1)$ code, which is optimal at this regime.

\begin{figure}[t]
\centering
\includegraphics[width = \onewidth]{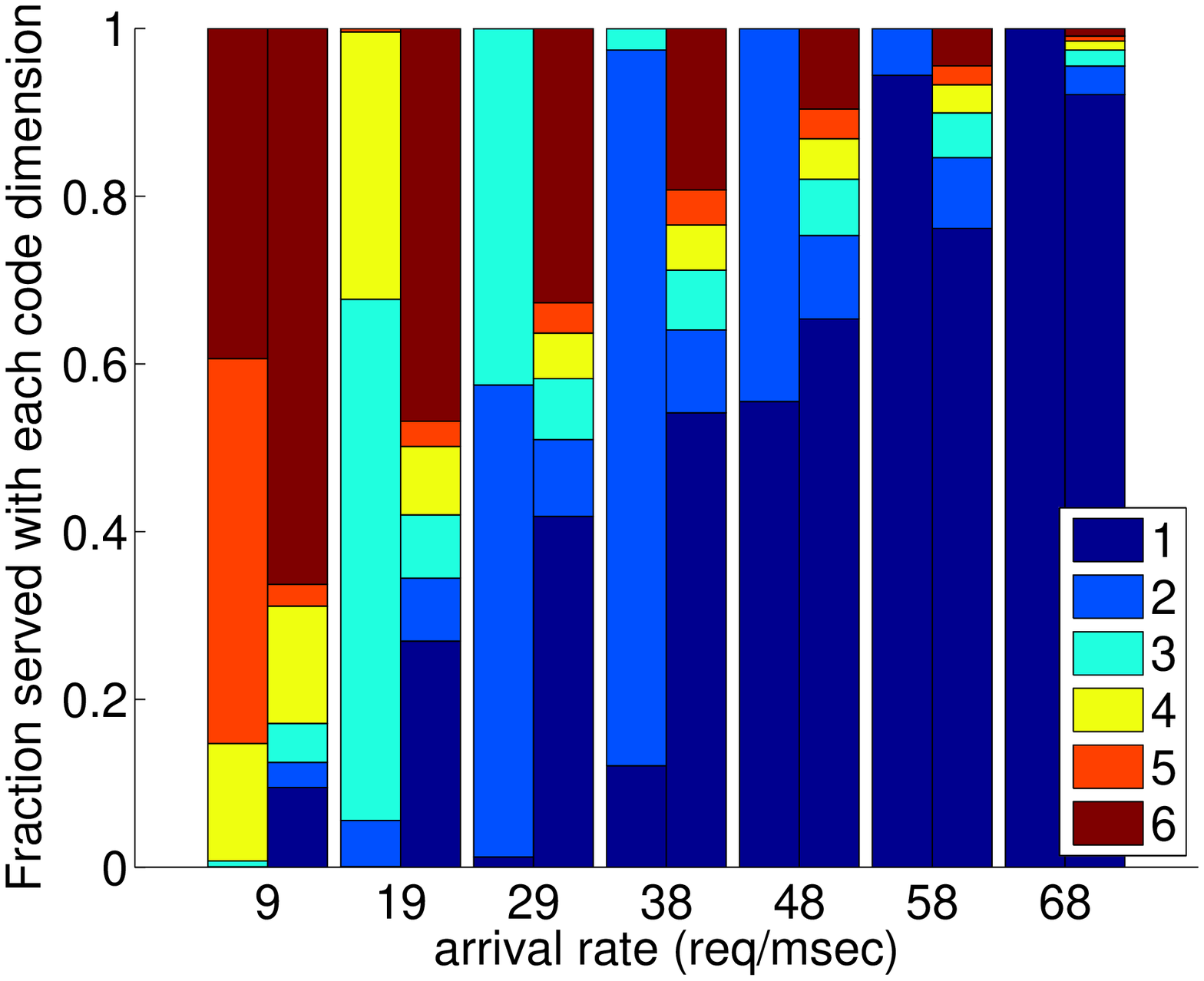}
\vspace{\shrinkbeforecaption}
\caption{Fraction of $k$. Left: \ourproposal, Right: Greedy}
\label{fig:read:composition}
\vspace{\shrinkaftercaption}
\end{figure}

\subsection{Behavior of the Adaptation Mechanisms}

\comment{
In Fig.\ref{fig:read:codeDim} and Fig.\ref{fig:read:codeLength} we plot the average code dimension $k$ and code length $n$ in \ourproposal and Greedy, as well as the code used by the static strategies that produce the best mean delay at different arrival rates. These figures again confirm that the adaptation algorithms of \ourproposal and Greedy are working as they are designed to: the average code dimension and length both match with the best static strategies quite well. Greedy is a bit too aggressive in choosing code dimension when compared with the best static strategy: the average code dimension of Greedy is always at least as large as that of the best static strategy. On the contrary,  \ourproposal's choice of code dimension turns out to be a better interpolation of the best static strategy.
}

\begin{figure}[t]
\centering
\includegraphics[width = \onewidth]{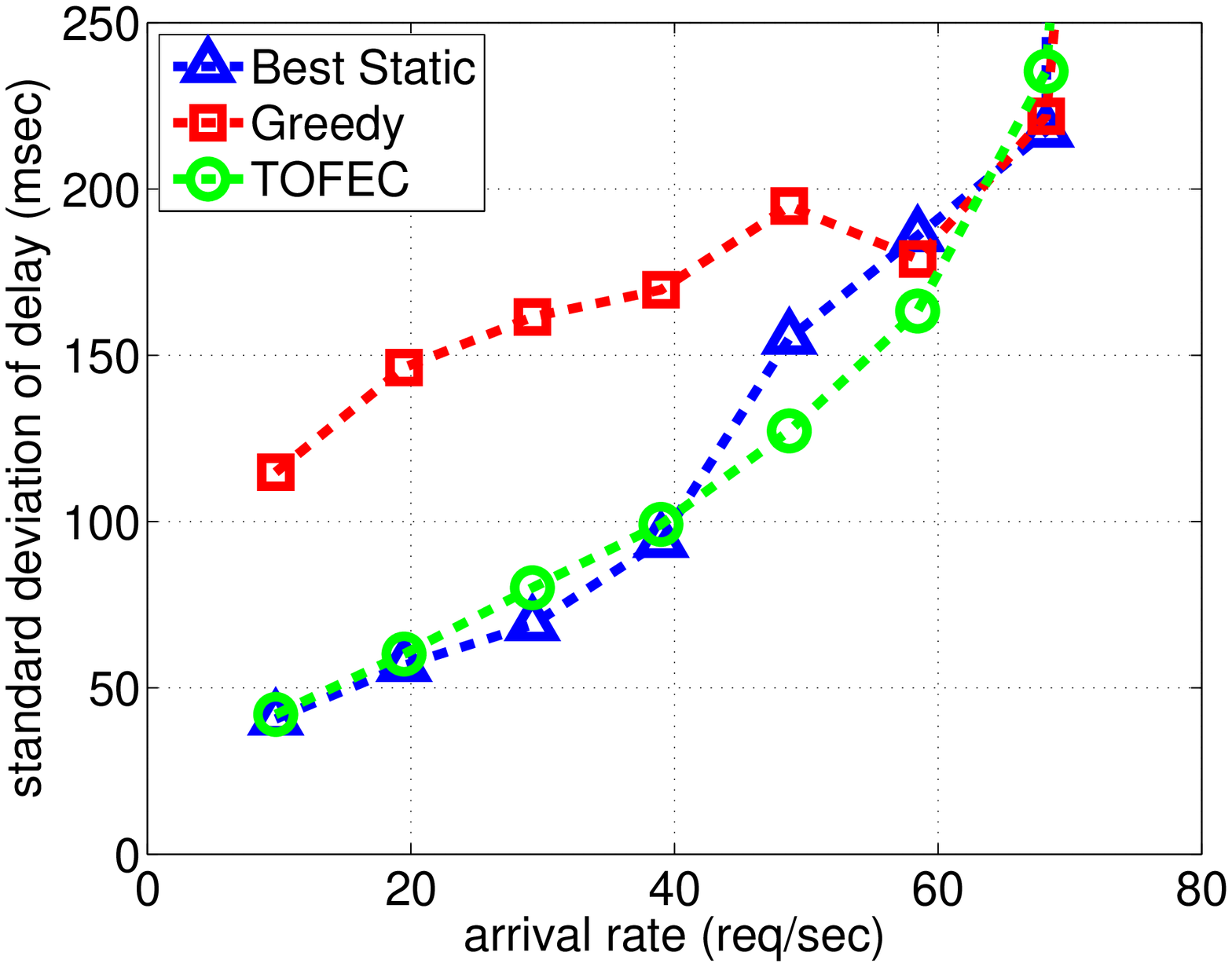}
\vspace{\shrinkbeforecaption}
\caption{Standard Deviation}
\label{fig:read:std}
\vspace{-10pt}
\end{figure}

When we look into the fraction of requests served by each choice of code, \ourproposal and Greedy
turn out to behave quite differently. In Fig.\ref{fig:read:composition} we plot the compositions of requests served by different code dimension $k$'s. 
At each arrival rate, the two bars represent \ourproposal and Greedy. For each bar, the colors
represent the fraction of requests served with code dimension
1 through 6, from bottom to top. \ourproposal's choice of $k$ demonstrates a high concentration around the optimal value: at all arrival rate, over 80\%  requests are served by 2 neighboring values of $k$. Moreover, as arrival rate varies from low to high, \ourproposal's choice of $k$ transitions quite smoothly as $(5,6) \rightarrow (3,4) \rightarrow  (2,3) \rightarrow (1,2)$ and eventually converges to a single value $1$ as workload approaches system capacity.
On the contrary, Greedy tends to round-robin across all possible choices of $k$ and majority of requests are served by either $k=1$ or $6$. So Greedy is effectively alternating between the two extremes of no chunking and very high chunking,  instead of staying around the optimal. Such ``all or nothing'' behavior results in $2\times$ to $3\times$ worse standard deviation as shown in Fig.\ref{fig:read:std}. So \ourproposal provides much better QoS guarantee.

\begin{figure}[t]
\centering
\includegraphics[width = \onewidth]{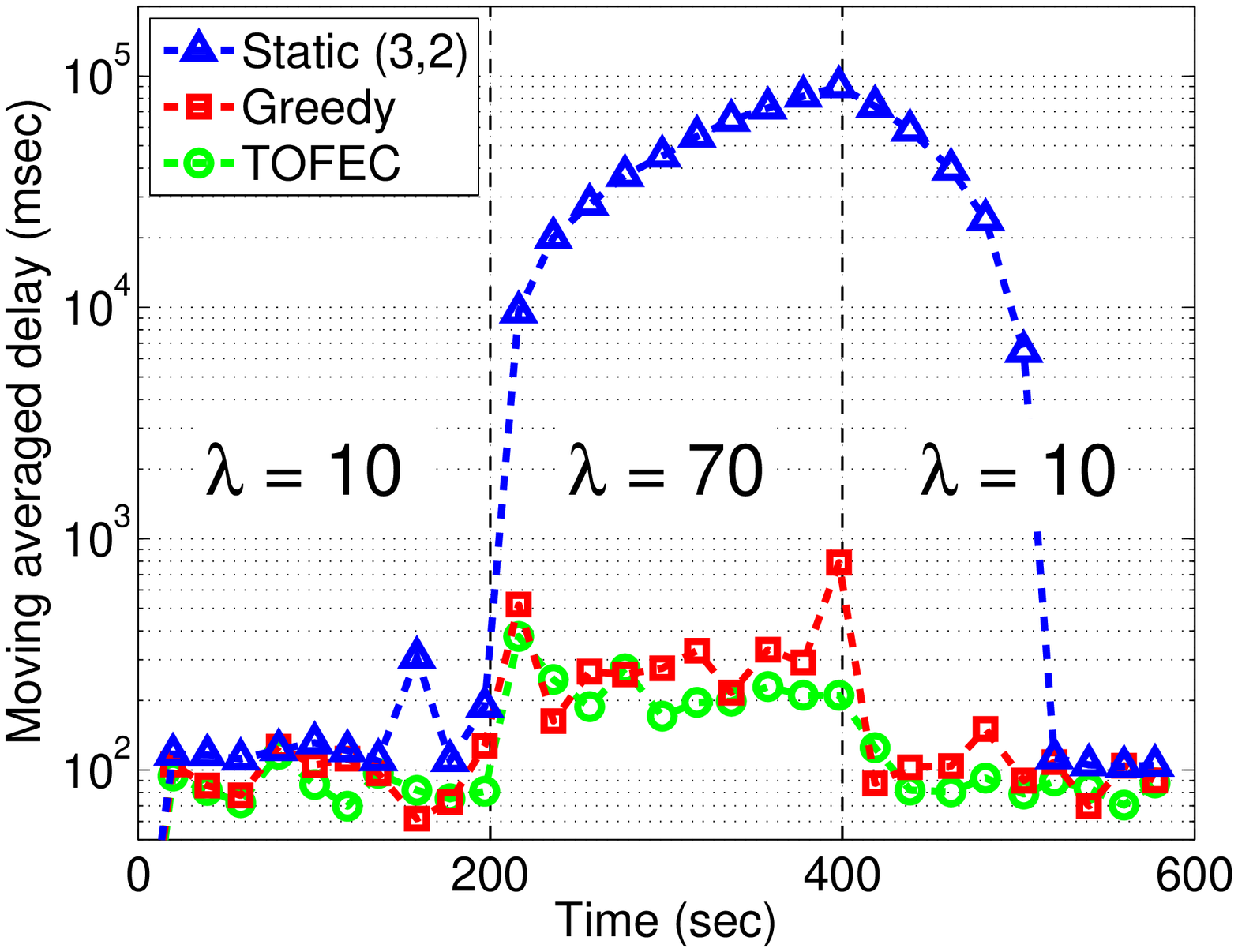}
\vspace{\shrinkbeforecaption}
\caption{Adaptation to changing workload}
\label{fig:read:changeArrival}
\vspace{\shrinkaftercaption}
\end{figure}

%\subsection{Adapting to Changing Workload}
We further examine how well the two adaptive strategies adjust to changes in workload. In Fig.\ref{fig:read:changeArrival} we plot the total delay experienced by requests arriving at different times within a 600-second period. The arrival rate is 10 request/second for the first and last 200 seconds, and 70 request/second for the middle 200 seconds. Both adaptive strategies turn out to be quite agile to changes in arrival rate and quickly converge to a good composition of codes that delivers optimal delays. On the contrary, the static strategy using $(3,2)$ code builds up a huge backlog during middle 200-second period and takes over 100 seconds to clean it up.

\section{Related Work}
\label{sec:related}

FEC in connection with multiple paths and/or multiple servers is a well investigated topic in the literature \cite{VickySharmaMPLOT,EminGabrielyanFEC,JohnByersAccessing,RSaadEvaluating}. However, there is very little attention devoted to the queueing delays. FEC in the context of network coding or coded scheduling has also been a popular topic from the perspectives of throughput (or network utility) maximization and throughput vs. service delay trade-offs \cite{Eryilmaz:2008:DTG:2263482.2273567,Yeownetworkcoding,Theodorosnetworkcoding, KozatScheduling}. Although some incorporate queuing delay analysis, the treatment is largely for broadcast wireless channels with quite different system characteristics and constraints.
FEC has also been extensively studied in the context of distributed storage from the points of high durability and availability while attaining high storage efficiency \cite{Dimakis:2010:NCD:1861840.1861868,Rodrigues_highavailability,Li:2010:TDR:1833515.1833884}. 

Authors of \cite{Longbocodeingincloud} conducted theoretical study of cloud storage systems using FEC in a similar fashion as we did in our work \cite{fastcloud}. Given that exact mathematical analysis of the general case is very difficult, authors of \cite{Longbocodeingincloud} considered a very simple case with a fixed code of $k=2$ tasks. Shah et al. \cite{MDS-queue} generalize the results from \cite{Longbocodeingincloud} to $k>2$. Both works rely on the assumption of exponential task delays, which hardly captures the reality. Therefore, some of their theoretical results cannot be applied in practice.
For example, under the assumption of exponential task delays, Shah et al. have proved that using larger $n$ will not reduce system capacity and will always improve delay, contradicting with simulation results using real-world measurements in \cite{fastcloud} and this paper. 
%They also included some simulation-based results for more non-exponential task delays resonate with our earlier findings in \cite{fastcloud}.

\section{Conclusion}
\label{sec:conclusion}
\ourproposal 's adaptation mechanism is the first technique for automatically adjusting the level of both chunking and redundancy for scalable key-value storage access using erasure codes and parallel connections. \ourproposal monitors the local backlog and dynamically adjust both the length and dimension of the erasure code to be used. To evaluate \ourproposal 's adaptation mechanism, we run simulations using real-world traces obtained on Amazon S3. We found that \ourproposal delivers the optima throughput-delay tradeoff and dramatically outperforms non-adaptive strategies and simple adaptive heuristics.

\bibliographystyle{IEEEtran}
\bibliography{PaperList}

\appendix
\section{Appendix}
\label{sec:appendix}

\begin{IEEEproof}
The objective of \eqref{eq:optimization} is a lower-bounded continuously differentiable function within the feasible region. Its value goes to $\infty$ as $(k,r)$ approaches the boundary of the feasible region. As a result, there exist at least one global optimal solution. At the global optimal, derivatives of the objective over $k_i$ and $r_i$ both equal to 0.
%, otherwise it cannot be a global optimal since the objective can be further reduced by following the decreasing direction of the gradient. 
Equating the partial derivatives of the objective over $k_i$ and $r_i$ to 0 can be rewritten into Eq.\ref{eq:opt:code} and Eq.\ref{eq:opt:normArrival}.

It is trivial to show that the left hand side of Eq.\ref{eq:opt:code} is a strictly increasing function of $k_i$ and the right hand side is a strictly increasing function of $r_i$ as long as $r_i\ge 1$. This implies that, $r_i$ is a strictly increasing function of $k_i$. The right hand side of Eq.\ref{eq:opt:normArrival} becomes some function $\pi_i(k_i)$ of $k_i$ by substituting $r_i$ with the solution from Eq.\ref{eq:opt:code}. It can be shown that $\pi_i$ is a strictly decreasing function.
Notice that Eq.\ref{eq:opt:normArrival} must be satisfied for all $i$ and the left hand side remains unchanged. Then
\begin{equation}
\pi_i(k_i) = \pi_j(k_j),~\forall i,j = 1,\cdots,m
,~\forall i,j.
\label{eq:pi:constant}
\end{equation}
Recall that $\pi_i$ and $\pi_j$ are strictly decreasing functions of $k_i$ and $k_j$, respectively. This means that there is a one-to-one mapping between any $k_i$ and $k_j$ at the optimal solutions, and $k_j$ is a strictly increasing function of $k_i$. 

Notice that for any given $\lambda$ and $\{p_i\}$ the left hand side of Eq.\ref{eq:opt:normArrival} is a strictly increasing function of $k_i$ if we replace all $k_j$'s and $r_j$'s with the solutions of Eq.\ref{eq:opt:code} and Eq.\ref{eq:pi:constant}. The right hand side of Eq.\ref{eq:opt:normArrival} is $\pi_i(k_i)$, which is a strictly decreasing function of $k_i$. As a result, these two functions can be equal for at most one value of $k_i$, i.e., Eq.\ref{eq:opt:code} and Eq.\ref{eq:opt:normArrival} have at most one solution. Since we have already proved the existence of a solution to these equations via the existence of global optimal, they have an unique solution.
\end{IEEEproof}

\end{document}